\documentclass{optica-article}

\journal{opticajournal} 

\articletype{Research Article}
\usepackage[utf8]{inputenc}
\usepackage{textcomp}
\usepackage{lineno}

\begin{document}

\title{Measurement of microwave polarization using two polarization orthogonal local microwave electric fields in a Rydberg atom-based mixer}

\author{Weibo Yin,\authormark{1,$\dag$} Jianan Zhang,\authormark{1,$\dag$} Fengdong Jia,\authormark{1,*} Yuhan Wang,\authormark{1} Yuxiang Wang,\authormark{1} Jianhai Hao,\authormark{1} Yue Cui,\authormark{1}  Ya Liu,\authormark{1,2} and Zhiping Zhong.\authormark{1,3}}

\address{\authormark{1}School of Physical Sciences, University of Chinese Academy of Sciences, Beijing 100049, China\\
\authormark{2}National Time Service Centre, Chinese Academy of Sciences, Xi’an 710600, People’s Republic of China\\
\authormark{3}CAS Center for Excellence in Topological Quantum Computation, University of Chinese Academy of Sciences, Beijing 100190, China\\
\authormark{$\dag$}The authors contributed equally to this work.}
\email{\authormark{*}fdjia@ucas.ac.cn} 

\begin{abstract*} 

We propose and demonstrate a novel method for measuring the polarization direction of a microwave electric field in a single measurement using a Rydberg atom-based mixer with two orthogonally polarized local microwave electric fields. Furthermore, introducing a weak static magnetic field enables the utilization of the Zeeman effect and exploitation of polarization asymmetry. This distinction allows for determining the polarization direction of the microwave field is \(\theta\) or \(180^\circ - \theta\) within the 0 to 180 degree range. This is the first real-time measurement of microwave polarization within 0 to 180 degrees, crucial for microwave sensing and information transmission. 

\end{abstract*}

\section{\textbf{Introduction}}

Quantum sensors achieve higher sensitivity and accuracy compared to traditional sensors\cite{degen2017RMP}. The application of quantum sensors in electromagnetic field measurements is particularly notable, especially in the measurement of microwave fields, which is crucial for wireless communication, radar systems, and astronomy\cite{sedlacek2012NP}. The measurement of microwave fields includes parameters such as electric field strength\cite{sedlacek2012NP,holloway2014IEEE,fan2015JPB,jing2020NP,anderson2021IEEE,jia2021PRA}, phase\cite{simons2019APL,simons2019IEEE,jia2021JPB,liu2022CPB}, and polarization direction\cite{sedlacek2013PRL,simons2019IEEE,jing2020NP,wang2023OE}. In particular, precise measurement of the polarization direction of microwave fields can provide information about the electromagnetic wave propagation environment, such as reflection, scattering, and absorption phenomena. Accurate measurement of the microwave polarization direction is essential for understanding and optimizing microwave transmission characteristics.

Recently, several research groups have made significant progress in using microwave electric field sensors to measure microwave polarization. In 2013, J. A. Sedlacek \textit{et al.} first utilized Rydberg-atom electromagnetically induced transparency (EIT) in a Rb atomic vapor cell to measure the polarization of a microwave field relative to the polarization of probe and coupling light, achieving an angular resolution of 0.5 degrees\cite{sedlacek2013PRL}. This method relies on the polarization of the light fields and requires complex modeling for accurate results. In 2022, Zhang \textit{et al}. calibrated the polarization direction of a microwave field using the EIT Autler–Townes splitting (EIT-AT) effect in a room-temperature Cs vapor cell, achieving an angular resolution of 1.64 degrees\cite{jing2020NP}. In 2023, Liu \textit{et al}. measured microwave polarization using Rydberg atoms' EIT-AT in cold atoms\cite{Liu2023AtomVectorMicrowaveElectricField}. For polarization measurement, Simons \textit{et al.} first demonstrated that a Rydberg atom-based mixer could determine the polarization of a microwave field\cite{simons2019IEEE}. Wang \textit{et al.} accurately measured microwave polarization using a Rydberg atom-based mixer\cite{wang2023OE}. Their results showed that the amplitude of beat note varied with the microwave field polarization in a 180-degree cycle, and within the linear region, a polarization resolution better than 0.5 degrees could be easily achieved. Interestingly, the measurement based on the mixer is not affected by the polarization of the light fields forming the Rydberg EIT. This method significantly simplifies the theoretical analysis and experimental system required for measuring microwave polarization using Rydberg atoms. Noted that, the above methods using Rydberg atom-based mixers to measure microwave polarization direction often require adjusting the polarization of the local microwave electric field to be rotated by 180 degrees to obtain the polarization direction of the microwave field, making the operation difficult. Furthermore, due to the symmetry of the projection of positive and negative polarization directions, it is impossible to distinguish whether the polarization direction of the microwave field is \(\theta\) or \(180^\circ - \theta\) within the 0-180 degree range in a single measurement. Typically, continuous measurements are used to distinguish based on the trend, making real-time high-resolution measurement of the microwave polarization direction challenging.

Building on the work of Wang \textit{et al.} \cite{wang2023OE}, this work introduces two orthogonally polarized local microwave electric fields in a Rydberg atom-based mixer to make the polarization measurement much simpler and convenient. We will demonstrate how to directly obtain the polarization direction of the signal field by comparing the beat signal strength of each local field with the signal field. We will further discuss the introduction of a weak magnetic field as the quantization axis, and attempt to distinguish between \(\theta\) and \(180^\circ - \theta\) within 0-180 degrees by comparing the changes in beat frequency signals with and without a magnetic field.

The structure of this paper is as follows: Section II theoretically demonstrates the relationship between the beat note signals output by the Rydberg atom-based mixer with two orthogonally polarized local microwave electric fields and the polarization direction of the microwave field to be measured. Section III introduces the experimental system and methods. Section IV presents the detailed experimental results of measuring the microwave polarization direction using two orthogonally polarized local microwave electric fields, as well as the impact of an additional static magnetic field on the polarization measurement. Section VI provides conclusions and a summary.

\section{Theoretical model}

\begin{figure}[ht!]
\centering\includegraphics[width=12cm]{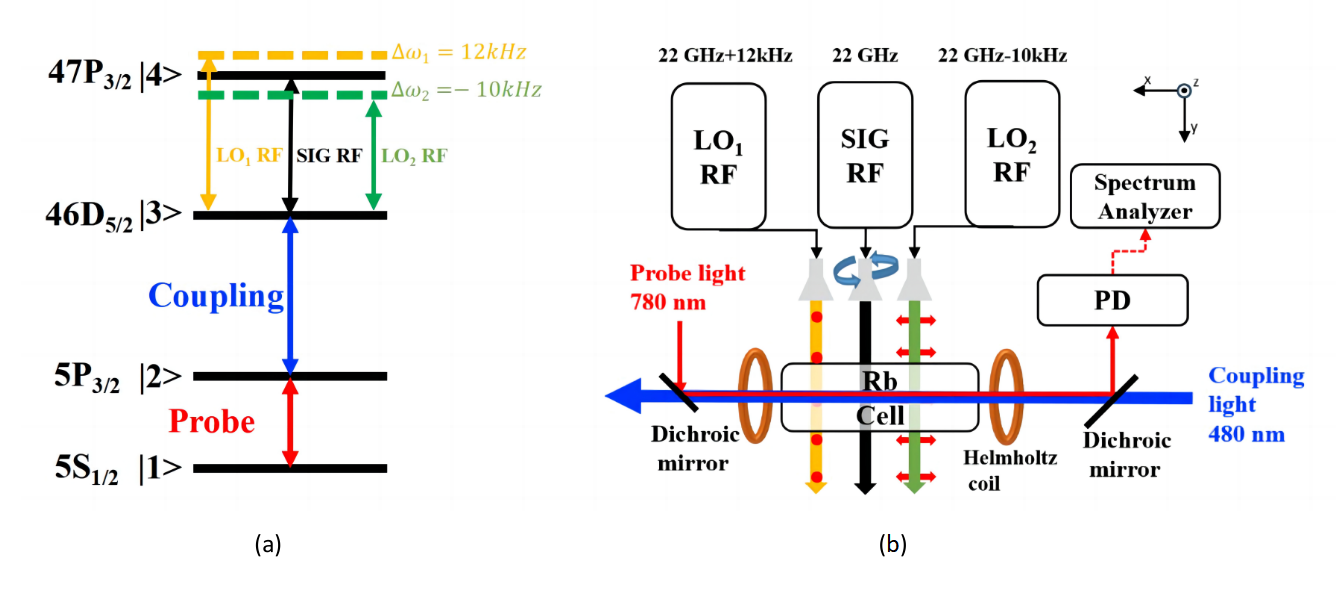}
\caption{(a) The energy levels involved in Rydberg EIT-AT and the mixer. The red arrow represents the probe light, which drives the transition between levels $\left| 1 \right\rangle$ and $\left| 2 \right\rangle$; the blue arrow represents the coupling light, which drives the transition between levels $\left| 2 \right\rangle$ and $\left| 3 \right\rangle$. The yellow, green, and black arrows represent the two local microwave fields and the signal microwave field, respectively, which drive the transition between levels $\left| 3 \right\rangle$ and $\left| 4 \right\rangle$. The frequency differences between the signal microwave field and the two local microwave electric fields are 12 kHz and -10 kHz, respectively. (b) The experimental setup for three-wave mixing to measure the polarization direction of the signal microwave electric field. The 780 nm probe light and 480 nm coupling light are overlapped and counter-propagated through two dichroic mirrors in a rubidium vapor cell. The two local microwave fields and the signal microwave field are radiated into the rubidium vapor through three rectangular horn antennas. The signal microwave electric field microwave horn is mounted on a rotating plate, which can rotate under electric control. The intensity signal of the probe light, obtained by a photodetector, is directly connected to a spectrum analyzer. The propagation direction of the 480 nm coupling light is defined as the positive x-axis, the y-axis as the propagation direction of the microwave fields, and the z-axis as perpendicular to both the x and y axes.}
\end{figure}

A Rydberg atom-based RF mixer utilizes Electromagnetically Induced Transparency (EIT) and Autler-Townes (AT) splitting phenomena in Rydberg atoms. When exposed to microwave fields, these atoms exhibit EIT peaks that split, causing modulation in the transmission of probe light. By applying two microwave fields with different frequencies, designated as the signal microwave and local microwave fields, this modulation directly reflects the interference between the two microwave signals. This enables the mixer to detect their beat note. The work of Simons \textit{et al.} \cite{simons2019IEEE} and Wang \textit{et al.} \cite{wang2023OE} demonstrates that when two linearly polarized RF microwave fields of different frequencies, \(E_{\mathrm{LO}}\) and \(E_{\mathrm{SIG}}\), coupled to the Rydberg energy levels, are applied to Rydberg atoms, the transmission of the probe light includes a beat notes with a frequency equal to the difference between the two microwave frequencies\cite{simons2019APL}.

\begin{equation}
T_{probe} \propto\left|E_{mod}\right|=\sqrt{E_{\mathrm{LO}}^2+E_{\mathrm{SIG}}^2+2 E_{\mathrm{LO}} E_{\mathrm{SIG}} \cos (\Delta \omega t+\Delta \phi)},
\end{equation}

where,\( T_{\text{probe}} \) represents the transmission of probe light. \( E_{\text{mod}} \) denotes the total electric field sensed by the atom, which includes contributions from the electric fields \( E_{\mathrm{LO}} \) of the local microwave and \( E_{\mathrm{SIG}} \) of the signal microwave. \( \Delta \omega \) signifies the frequency difference between the signal and local microwaves, defined as \( \Delta \omega = \omega_{\mathrm{LO}} - \omega_{\mathrm{SIG}} \), where \( \omega_{\mathrm{SIG}} \) and \( \omega_{\mathrm{LO}} \) are their angular frequencies. \( \Delta \phi \) represents the phase difference between these microwaves, defined as \( \Delta \phi = \phi_{\mathrm{LO}} - \phi_{\mathrm{SIG}} \), where \( \phi_{\mathrm{SIG}} \) and \( \phi_{\mathrm{LO}} \) denote their respective phases. When the angle between the polarization directions of the two linearly polarized microwave fields is $\theta$, the amplitude of the beat note is \cite{simons2019APL},

\begin{equation}
A_{beat} \propto\sqrt{E_{\mathrm{LO}}^2+E_{\mathrm{SIG}}^2+2 E_{\mathrm{LO}} E_{\mathrm{SIG}} \cos \theta} = \left|E_{\mathrm{SIG}} + E_{\mathrm{LO}} \cos \theta\right|,
\end{equation}

where, \( A_{\text{beat}} \) represents the amplitude of the beat frequency signal. \( \theta \) is defined as the angle between the polarizations of the signal microwave and local microwave fields. We introduced three microwave fields with frequencies close to each other, coupled to the same pair of Rydberg energy levels. We define two local microwave electric fields and one signal microwave electric field as follows:

\begin{equation}
E_1 = E_{\mathrm{LO1}} \cos \left(\omega_{\mathrm{LO1}} t + \phi_{\mathrm{LO1}}\right), \quad E_2 = E_{\mathrm{LO2}} \cos \left(\omega_{\mathrm{LO2}} t + \phi_{\mathrm{LO2}}\right),
\quad E_3 = E_{\mathrm{SIG}} \cos \left(\omega_{\mathrm{SIG}} t + \phi_{\mathrm{SIG}}\right).
\end{equation}

The frequency differences between the signal microwave electric field and the two local microwave electric fields are defined as,

\begin{equation}
\Delta \omega_{1} = \omega_{\mathrm{LO1}} - \omega_{\mathrm{SIG}} , \quad \Delta \omega_{2} = \omega_{\mathrm{LO2}} - \omega_{\mathrm{SIG}}.
\end{equation}

The polarization directions of the two local microwave electric fields, $E_{1}$ and $E_{2}$ are orthogonal, with the polarization direction of $E_{1}$ along the z-axis and $E_{2}$ along the x-axis. Orthogonally polarized microwave fields do not produce beat notes in principle. Therefore, the signal microwave electric field only mixes with the projection of its vector onto the vector of a local microwave electric field. Thus, the signal microwave electric field can be orthogonally decomposed into components parallel to the local microwave electric field 1 and parallel to the local microwave electric field 2. If the angle between the polarization direction of the signal microwave field and the z-axis is $\theta$, 
\begin{equation}
   \mathbf{E}_{\mathrm{SIG}} = \frac{E_{\mathrm{SIG}} \cos \theta}{|E_{\mathrm{LO1}}|} \mathbf{E}_{\mathrm{LO1}} + \frac{E_{\mathrm{SIG}} \sin \theta}{|E_{\mathrm{LO2}}|} \mathbf{E}_{\mathrm{LO2}}.
\end{equation}

And the amplitudes of the two beat notes formed by the signal microwave field and the two local microwave electric fields are respectively:

\begin{equation}
A_{beat1} \propto \left|E_{\mathrm{SIG}} + E_{\mathrm{LO1}} \cos \theta\right|,
\label{beat1}
\end{equation}

\begin{equation}
A_{beat2} \propto \left|E_{\mathrm{SIG}} + E_{\mathrm{LO2}} \sin \theta\right|.
\label{beat2}
\end{equation}

Because \( E_{\mathrm{SIG}} \) is much smaller than \( E_{\mathrm{LO}} \), the ratio of the two beat frequencies is given by:

\begin{equation}
\frac{A_{\text{beat2}}}{A_{\text{beat1}}} \approx \left| \frac{E_{\mathrm{LO2}} \sin \theta}{E_{\mathrm{LO1}} \cos \theta} \right| .
\end{equation}

If \( E_{\mathrm{LO1}} = E_{\mathrm{LO2}} \), then:

\begin{equation}
\frac{A_{\text{beat2}}}{A_{\text{beat1}}} \approx |\tan \theta|.
\label{tan}
\end{equation}

Figure 2 shows the curves of the amplitudes of the two beat notes formed by the signal microwave electric field and the two orthogonally polarized local microwave electric fields, plotted according to Eq.\eqref{beat1} and \eqref{beat2}, as a function of the polarization direction of the signal microwave electric field. The inset in the upper right corner illustrates the ratio of the two beat notes. It can be observed that the ratio obtained from a single measurement of the two beat notes directly provides the polarization direction of the signal microwave electric field, thus avoiding the issue of variations in single beat notes intensity caused by changes in signal microwave electric field strength as Eq.\eqref{tan}.

\begin{figure}[ht!]
\centering\includegraphics[width=12cm]{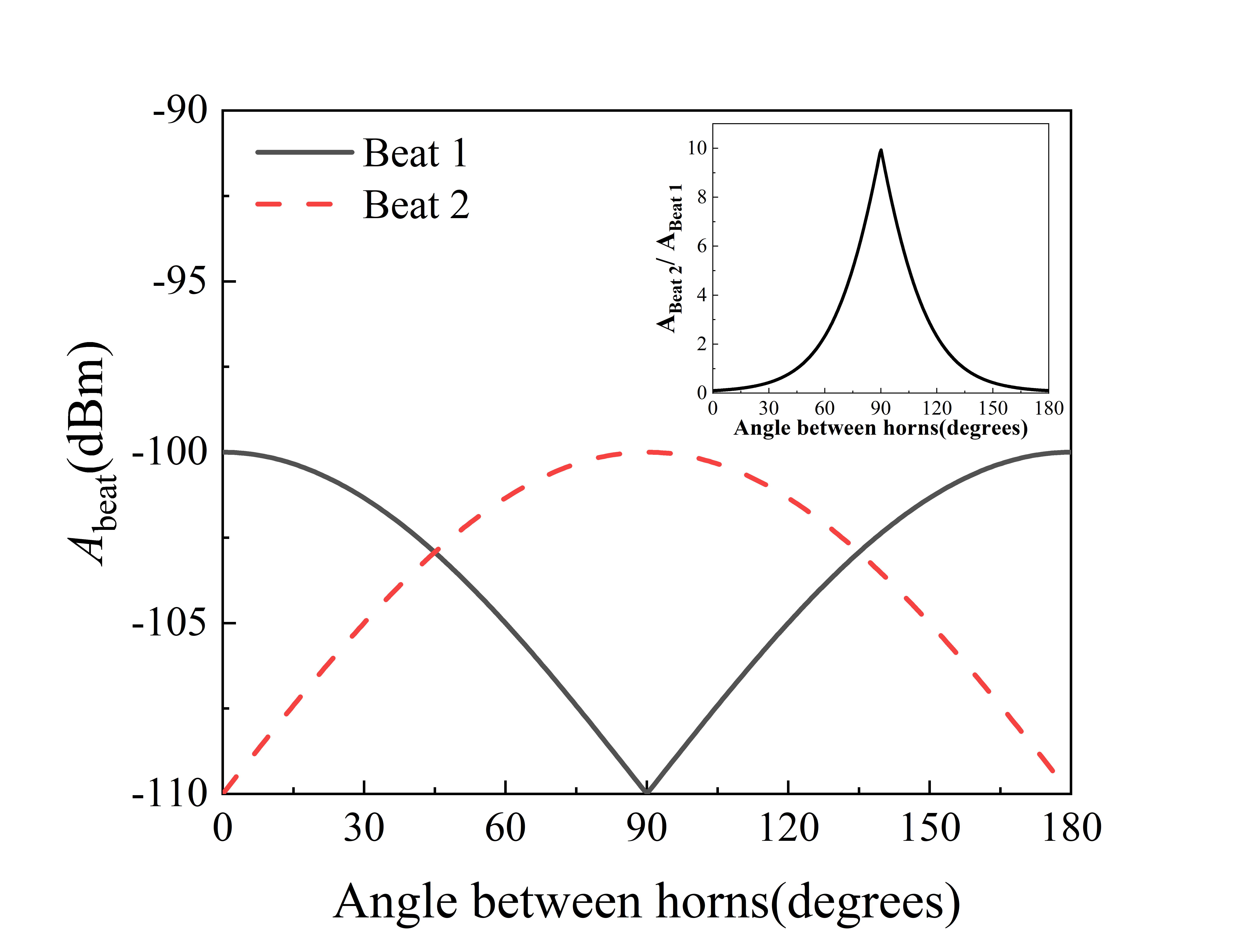}
\caption{Theoretical plots of the amplitudes of the two beat notes as functions of the polarization direction of the signal microwave electric field. The curves depict the variations in the amplitudes of the beat notes with the polarization angle as described by Eq.\eqref{beat1}  and \eqref{beat2}. The inset in the upper right corner shows the relationship between the ratio of the two amplitudes of the beat notes and the polarization direction of the signal microwave electric field.}
\end{figure}

The references \cite{li2023OE38165--38178,shi2023OE36255--36262,schlossberger2024PRAA} provide detailed studies on Rydberg EIT-AT splitting under a magnetic field. The reference \cite{schlossberger2024zeeman} shows that a quantized axis magnetic field and the polarization of EIT light can alter the distribution of atoms across different Zeeman sub-energy levels and shift the EIT spectra. For example, when the linear polarization of EIT light is perpendicular to a quantized axis magnetic field, it causes the atomic population to occupy the two states with the maximum absolute value of |mF|. Sigma+ circular polarization of EIT light causes the atomic population to reach the state with the maximum positive mF, while sigma- circular polarization results in the population occupying the state with the maximum negative mF. Thus, when the polarization of the microwave field and the quantization axis magnetic field are at a certain angle, if this angle is less than 80 degrees, there will be a greater population of atoms in positive mF states. Conversely, if the angle exceeds 80 degrees, the atoms are more likely to occupy negative mF states. This leads to asymmetric peak changes in the EIT spectrum, and may cause the maximum beat signal to deviate from being parallel to the signal field polarization and LO1. Consequently, this deviation allows for the qualitative judgment of angles \(\theta\) and \(180^\circ - \theta\) within the 0-180 degree range, achieving polarization resolution of \(\theta\) and \(180^\circ - \theta\).

\section{Experimental setup}

As shown in Figure 1(a), our experiment with the Rydberg atom-based mixer involves four energy levels in \(^{87}\mathrm{Rb}\) atoms: \(5S_{1/2}(F = 2)\), \(5P_{3/2}(F = 3)\), \(46D_{5/2}(F = 4)\), and \(47P_{3/2}(F = 3)\). The experimental setup for the Rydberg atom-based mixer we used is shown in Figure 1(b). The probe light is generated by a tunable diode laser (DL100, Toptica), with its frequency locked to the \(5S_{1/2}(F = 2) \rightarrow 5P_{3/2}(F = 3)\) transition of the \(^{87}\mathrm{Rb}\) atom. The 480 nm coupling light is produced by a frequency-doubled diode laser (TA-SHG-Pro, Toptica) and is frequency-locked to the \(5P_{3/2}(F = 3) \rightarrow 46D_{5/2}(F = 4)\) transition of the \(^{87}\mathrm{Rb}\) atom using Zeeman modulation EIT methods\cite{jia2020AO}. The probe light has an intensity of 60 $\mu$W and a diameter of 800 $\mu$m, while the coupling light has an intensity of 80 mW and a diameter of 900 $\mu$m.

The probe and coupling lights are overlapped along the central axis of the Rb atom vapor cell and propagate in opposite directions, forming Rydberg EIT. A photodetector(PDA8A2, Thorlabs) is used to receive the probe light transmitted through the Rb vapor cell, and an oscilloscope is employed to record the EIT and EIT-AT split spectra. In the experiment, three independent microwave antennas are used to irradiate the atomic cell with microwaves from the two local microwave electric fields and the signal microwave electric field, forming beat notes. The output from the photodetector is fed to a spectrum analyzer, where the two beat notes with different frequency are directly measured simultaneously. A pair of Helmholtz coils is used to generate a uniform static magnetic field in the positive x-axis direction, which is employed to cancel the horizontal component of the Earth's magnetic field and provide a static magnetic field.

Based on the principle of measuring microwave polarization using a mixer \cite{wang2023OE}, we adopted the traditional excitation scheme of $5S-5P-46D-47P$. To avoid the interference of the polarization directions of the probe light and coupling light on the beat notes after introducing a static magnetic field, we set the polarization states of both the probe light and coupling light to the same circular polarization\cite{wang2023OE}. Due to the selection rules for transitions, the circularly polarized probe and coupling light not only change the population of the energy levels but also cause some atoms in certain states not to interact with the microwaves\cite{sedlacek2013PRL}. As a result, the Rydberg EIT-AT spectrum is no longer a double-peak structure but a three-peak structure that is a mixture of the three-level EIT spectrum and the four-level EIT-AT split spectrum. To avoid the influence of atoms that only participate in the three-level EIT process and thus do not interact with the microwaves on the measurement, we locked the probe light frequency to the outer side of the EIT-AT split peaks, with a detuning of approximately 5 MHz from the resonance position.

In the experiment, the settings for the three microwave fields are as follows. First, to meet the far-field conditions, we placed the three microwave horn antennas at a distance from the Rb vapor cell. The three microwave antennas are placed parallel to each other; the microwave antennas on both sides output local microwave electric field 1 and local microwave electric field 2, while the central microwave antenna outputs the signal microwave electric field. The frequency settings for the three microwave fields are as follows: the signal microwave electric field frequency is set to resonate with the $46 D_{5/2}-47P_{3/2}$ transition, at 22.067 GHz. The frequency of local microwave electric field 1 is set to 22.067 GHz + 12 kHz, and the frequency of local microwave electric field 2 is set to 22.067 GHz - 10 kHz, thus forming beat notes of 10 kHz and 12 kHz with the signal microwave electric field, respectively. Both local microwave electric fields are linearly polarized. In the experiment, we fixed the intensity and polarization direction of the two local microwave electric fields, with the polarization direction of local microwave electric field 1 along the z-axis and the polarization direction of local microwave electric field 2 along the x-axis. By carefully adjusting the directions of the rectangular horn antennas of the two local microwave electric fields, we achieved mutual orthogonality of their polarizations. This was achieved precisely by observing the 22 kHz beat notes generated from the two local microwave electric fields, ensuring minimal signal when the polarizations are orthogonal. The signal microwave electric field is also irradiated onto the atomic cell through a linear rectangular horn antenna, which is mounted on an electric rotating plate to change the polarization direction. The motorized rotating stage, controlled by a stepper motor, can rotate continuously within a range of 360 degrees with a stepping accuracy of 0.01 degrees. In the experiment, the polarization direction of the signal microwave electric field can change continuously within a 360-degree range. Initially, we adjusted the polarization directions and amplitudes of  two local microwave electric field microwaves. Then, we adjusted the output power of the three microwave sources to ensure that the maximum and minimum values of the beat notes at 12 kHz and 10 kHz generated by the signal microwave field with the two local microwave fields were approximately equal. The output power of the microwave source for local microwave electric field 1 is 7 dBm, for local microwave electric field 2 is 9 dBm, and for the signal microwave electric field is 1 dBm.

According to the theoretical analysis in Section II, the intensities of these two beat notes are related to the projections of the signal microwave electric field onto the polarization directions of the two orthogonally polarized local microwave electric fields. By measuring the amplitudes of the two beat notes in a single measurement and analyzing their relative magnitudes, it is possible to achieve high-resolution real-time measurement of the polarization direction of the signal microwave electric field within the range of 0 to 90 degrees. However, due to the symmetry of the projections of the positive and negative polarization directions of the signal microwave electric field onto the local microwave electric fields, the aforementioned measurement method and previous mixer measurement schemes both struggle to distinguish whether the polarization direction of the microwave electric field is \(\theta\) or \(180^\circ - \theta\) within the range of 0-180 degrees. We utilized a pair of Helmholtz coils to add an additional static magnetic field of 4 G along the positive z-axis near the rubidium vapor cell, aiming to distinguish whether the polarization direction of the signal microwave electric field is \(\theta\) or \(180^\circ - \theta\).

\section{Results and discussion}

\begin{figure}[ht!]
\centering\includegraphics[width=12cm]{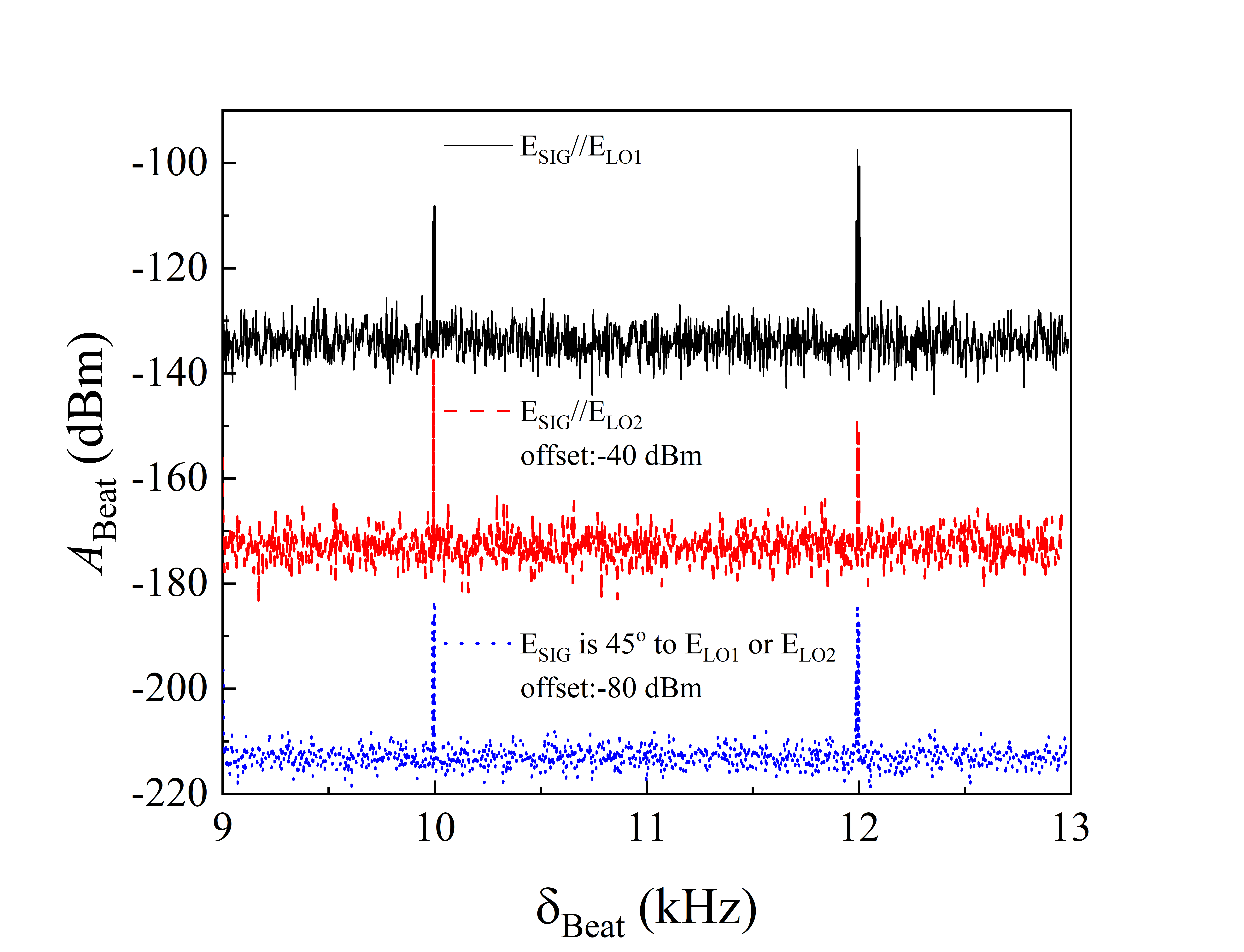}
\caption{Typical beat notes generated by the two local microwave electric fields when the signal microwave electric field has different polarization directions. (Top) The polarization direction of the signal microwave electric field is the same as that of the local microwave field 1. (Middle) The polarization direction of the signal microwave electric field is the same as that of the local microwave field 2. (Bottom) The polarization direction of the signal microwave electric field forms a 45-degree angle with the polarization directions of both local microwave fields 1 and 2.}
\end{figure}

We will first experimentally demonstrate the changes in the beat frequency signals of the signal field and two local fields when changing the polarization direction of the signal field. Figure 3 illustrates the intensity variations of the beat notes generated by the two local microwave electric fields when the signal microwave electric field has different polarization directions. Specifically, as shown in the top part of Figure 3, when the polarization direction of the signal microwave electric field aligns with that of the local microwave field 1, the beat notes (12 kHz) reaches its maximum. At this time, the polarization direction of the signal microwave electric field is perpendicular to that of the local microwave electric field 2, resulting in a minimum beat notes (10 kHz). Note that the beat notes is not zero due to the circular polarization of the probe and coupling light\cite{wang2023OE}. Similarly, as shown in the middle part of Figure 3, when the polarization direction of the signal microwave electric field aligns with that of the local microwave field 2, the 10 kHz beat notes reaches its maximum, while the 12 kHz beat notes is at its minimum. The bottom part of Figure 3 shows that when the signal microwave electric field forms a 45-degree angle with the polarization directions of both local microwave electric fields 1 and 2, the amplitude of the two beat notes is nearly identical.

\begin{figure}[ht!]
\centering\includegraphics[width=12cm]{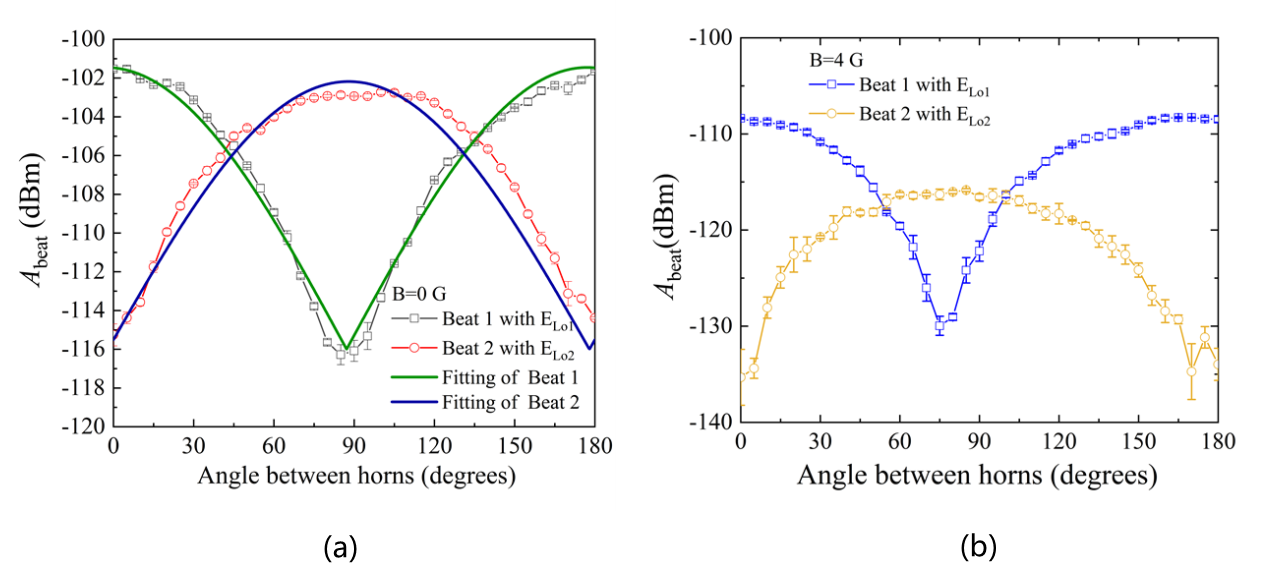}
\caption{(a) shows the relationship between the amplitudes of two beat notes (12 kHz in black and 10 kHz in red) and the polarization angle of the signal microwave electric field in the absence of a magnetic field. The horizontal axis represents the rotation angle of the signal microwave electric field, and the vertical axis represents the amplitudes of the beat notes. The green and blue curves are the fitting results based on Eq.\eqref{beat1}  and \eqref{beat2}. (b) shows the relationship between the amplitudes of two beat notes (12 kHz in blue and 10 kHz in orange) and the polarization angle of the signal microwave electric field when B=4 G.}
\end{figure}

We investigated the variation patterns of the two beat notes when changing the polarization of the signal microwave electric field within the 0-180 degree range. Specifically, we used an electric rotating plate to change the polarization direction of the signal microwave electric field by 5 degrees each time, then recorded the amplitudes of the 12 kHz and 10 kHz beat notes on a spectrum analyzer. The relationship between the amplitude of the beat notes and the polarization direction of the signal microwave electric field is shown in Figure 4(a). The black curve represents the 12 kHz beat notes formed by the signal microwave electric field and local microwave field 1, while the red curve represents the 10 kHz beat notes formed by the signal microwave electric field and local microwave field 2. It is clear that within the 180-degree period, the amplitudes of the two beat notes vary with the polarization angle of the signal microwave electric field. When the signal microwave electric field angle is 90 degrees, the polarization direction of the signal microwave electric field is parallel to that of the local microwave electric field 2 and perpendicular to that of the local microwave electric field 1, resulting in a maximum amplitude for the 10 kHz beat notes and a minimum amplitude for the 12 kHz beat notes. When the signal microwave electric field is oriented at 0 or 180 degrees, its polarization direction is parallel to local microwave electric field 1 and perpendicular to local microwave electric field 2, resulting in a maximum amplitude for the 12 kHz beat notes and a minimum amplitude for the 10 kHz beat notes. The blue and green curves in Figure 4 represent the fitting results of the experimental data based on Eq.\eqref{beat1}  and \eqref{beat2}. The fitting results show good agreement between experimental and theoretical expectations. Subsequently, we investigated the ratio of the beat note amplitudes generated by local microwave electric fields at 10 kHz and 12 kHz with respect to the polarization direction of the signal microwave electric field. The experimental and theoretical results are shown in Figure 5 as black data points and a blue solid line. In practical experiments, factors such as microwave reflections and external interferences complicate the interpretation of the ratio of the two beat notess using a simple formula. As a result, there is some discrepancy between experimental data and the theoretical value in Eq.\eqref{tan}.

With this measurement approach, there is no longer a need to adjust the angle of the signal (or local) horn antennas multiple times for measurements. By using fixed-angle local microwave electric fields, a single measurement provides the ratio of the beat note amplitudes, enabling the determination of the polarization direction of the signal microwave field with high resolution in real time. Moreover, this method of projecting the signal microwave electric field polarization onto two orthogonal directions avoids issues arising from variations in signal microwave electric field intensity that affect the amplitude of a single beat note.

\begin{figure}[ht!]
\centering\includegraphics[width=12cm]{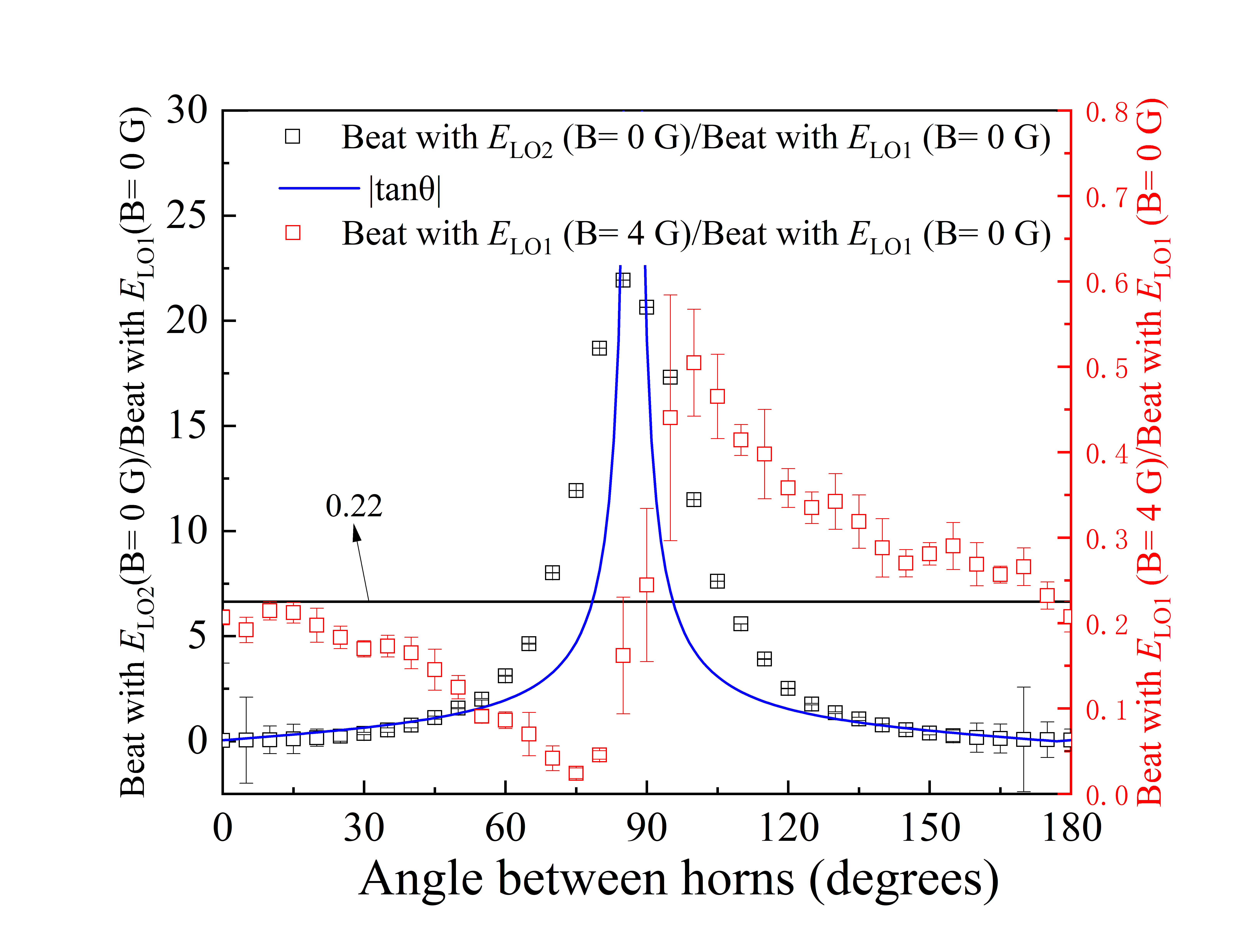}
\caption{The influence of a static magnetic field on the beat note signals. The black points illustrate the relationship between the ratio of the beat note amplitudes at 12 kHz and 10 kHz, without an additional magnetic field, and the polarization angle of the signal microwave field. The blue solid line represents \(|\tan \theta|\). The red points show how the ratio of the beat note amplitudes at 12 kHz, under conditions of both \( B = 4 \, \text{G} \) and no magnetic field, varies with the polarization angle of the signal microwave field. }
\end{figure}

From Figure 4(a) it is apparent that it is not possible to distinguish between polarization angles of X and-180-X degrees. To differentiate between these two angles, we introduced a background magnetic field of 4 G. Specifically, using a pair of Helmholtz coils, we applied a magnetic field of 4 G along the positive z-axis at the Rb cell. Continuing with the setup, we rotated the polarization direction of the signal microwave electric field every 5 degrees using an electric rotating plate. Subsequently, we read the corresponding beat note amplitudes at 12 kHz and 10 kHz on the spectrum analyzer, as shown in Figure 4(b). The magnetic field induced changes in the EIT spectrum, causing variations in the transmission rate of the probe light and thus resulting in a decrease in the amplitude of the beat notes\cite{jia2021JPB}. Figure 4(b) illustrates that, compared to no magnetic field, the application of a 4 G magnetic field shifted the minimum point of the 12 kHz beat note amplitude generated by local microwave field 1 by 10 degrees.

In Figure 5 we show how the ratio of the 12 kHz beat note amplitude between \(B = 4\) G and \(B = 0\) varies with the polarization angle of the signal microwave field, as shown in Figure 5. Using measurements from switching the magnetic field, when the ratio is less than 0.22, the signal microwave field is within the range of 0 to 90 degrees, and when the ratio is greater than 0.22, the signal microwave field is within the range of 90 to 180 degrees. This ratio depends on the magnetic field magnitude and the shift in the probe light frequency, but it is consistent once the system parameters are fixed. Therefore, through two measurements with switched magnetic fields, we achieve high-resolution, real-time polarization angle measurements of the signal microwave electric field within the 0-180 degree range.

Recently, a new study by Elgee \textit{et al}.\cite{Elgee2024three-dimensional} has been published, which explores similar aspects of microwave polarization measurements using Rydberg atoms. While their approach shares some similarities with ours, such as the use of orthogonally polarized local microwave fields, our method introduces several key innovations. Specifically, our technique allows for real-time polarization angle measurements with high resolution by capturing beat notes from both local microwave fields simultaneously in a single measurement and obtaining the ratio of the two beat signals. Additionally, by applying a weak static magnetic field, we can uniquely distinguish microwave electric field polarizations at \(\theta\) and \(180^\circ - \theta\), which has not been addressed in the recent study. These distinctions not only enhance the accuracy of our measurements but also expand the applicability of Rydberg atom-based sensors in practical scenarios.

\section{Conclusion}
In this paper, we present a novel method for measuring the polarization direction of a microwave electric field using two orthogonally polarized local microwave fields. Our method allows for the capture of the beat notes of the signal microwave electric field with both local microwave electric fields in a single measurement, and the ratio of these two beat notes allows for the accurate determination of the polarization direction of the signal microwave field. This enables high-resolution, real-time polarization angle measurements of the microwave polarization. Additionally, by introducing a weak static magnetic field of 4 G, we break the symmetry of atomic energy levels and beat note signals, achieving differentiation between microwave electric field polarizations at \(\theta\) or \(180^\circ - \theta\) within the 0-180 degree range. This work demonstrates, for the first time, the direct real-time measurement of microwave polarization within the 0-180 degree range, which holds significant implications for microwave sensing and information transmission. This breakthrough opens new avenues for enhancing the performance and reliability of microwave-based technologies. This measurement scheme paves the way for developing miniaturized, high-precision Rydberg atom-based microwave polarization analyzers for real-time measurements.

\begin{backmatter}
\bmsection{Funding}
Beijing Municipal Natural Science Foundation (1212014); National Natural Science Foundation of China(11604334); Fundamental Research Funds for the Central Universities; Youth Innovation Promotion Association of the Chinese Academy of Sciences.

\bmsection{Acknowledgments}

We would like to express our deepest gratitude to Professor Thomas F. Gallagher of University of Virginia for his invaluable guidance and insightful discussions, which significantly contributed to the finish of this research. We also extend our heartfelt thanks to Professors Yuechun Jiao, Linjie Zhang, and Jie Ma of Shanxi University; Dr. Kaiyu Liao and Professor Hui Yan of South China Normal University; and Dr. Fei Meng of the National Institute of Metrology, China, for their helpful discussions and support.

\bmsection{Disclosures}
 The authors declare no conflicts of interest.

\bmsection{Data Availability Statement}
Data underlying the results presented in this paper are not publicly available at this time but may be obtained from the authors upon reasonable request.

\end{backmatter}

\bibliography{references}

\begin{thebibliography}{10}
\newcommand{\enquote}[1]{``#1''}

\bibitem{degen2017RMP}
C.~L. Degen, F.~Reinhard, and P.~Cappellaro, \enquote{Quantum sensing,} {\protect\JournalTitle{Reviews of modern physics}} \textbf{89}, 035002 (2017).

\bibitem{sedlacek2012NP}
J.~A. Sedlacek, A.~Schwettmann, H.~K{\"u}bler, \emph{et~al.}, \enquote{Microwave electrometry with rydberg atoms in a vapour cell using bright atomic resonances,} {\protect\JournalTitle{Nature physics}} \textbf{8}, 819--824 (2012).

\bibitem{holloway2014IEEE}
C.~L. Holloway, J.~A. Gordon, S.~Jefferts, \emph{et~al.}, \enquote{Broadband rydberg atom-based electric-field probe for si-traceable, self-calibrated measurements,} {\protect\JournalTitle{IEEE Transactions on Antennas and Propagation}} \textbf{62}, 6169--6182 (2014).

\bibitem{fan2015JPB}
H.~Fan, S.~Kumar, J.~Sedlacek, \emph{et~al.}, \enquote{Atom based rf electric field sensing,} {\protect\JournalTitle{Journal of Physics B: Atomic, Molecular and Optical Physics}} \textbf{48}, 202001 (2015).

\bibitem{jing2020NP}
M.~Jing, Y.~Hu, J.~Ma, \emph{et~al.}, \enquote{Atomic superheterodyne receiver based on microwave-dressed rydberg spectroscopy,} {\protect\JournalTitle{Nature Physics}} \textbf{16}, 911--915 (2020).

\bibitem{anderson2021IEEE}
D.~A. Anderson, R.~E. Sapiro, and G.~Raithel, \enquote{A self-calibrated si-traceable rydberg atom-based radio frequency electric field probe and measurement instrument,} {\protect\JournalTitle{IEEE Transactions on Antennas and Propagation}} \textbf{69}, 5931--5941 (2021).

\bibitem{jia2021PRA}
F.-D. Jia, X.-B. Liu, J.~Mei, \emph{et~al.}, \enquote{Span shift and extension of quantum microwave electrometry with rydberg atoms dressed by an auxiliary microwave field,} {\protect\JournalTitle{Physical Review A}} \textbf{103}, 063113 (2021).

\bibitem{simons2019APL}
M.~T. Simons, A.~H. Haddab, J.~A. Gordon, and C.~L. Holloway, \enquote{A rydberg atom-based mixer: Measuring the phase of a radio frequency wave,} {\protect\JournalTitle{Applied Physics Letters}} \textbf{114} (2019).

\bibitem{simons2019IEEE}
M.~T. Simons, A.~H. Haddab, J.~A. Gordon, \emph{et~al.}, \enquote{Embedding a rydberg atom-based sensor into an antenna for phase and amplitude detection of radio-frequency fields and modulated signals,} {\protect\JournalTitle{IEEE access}} \textbf{7}, 164975--164985 (2019).

\bibitem{jia2021JPB}
F.-D. Jia, H.-Y. Zhang, X.-B. Liu, \emph{et~al.}, \enquote{Transfer phase of microwave to beat amplitude in a rydberg atom-based mixer by zeeman modulation,} {\protect\JournalTitle{Journal of Physics B: Atomic, Molecular and Optical Physics}} \textbf{54}, 165501 (2021).

\bibitem{liu2022CPB}
X.-B. Liu, F.-D. Jia, H.-Y. Zhang, \emph{et~al.}, \enquote{An all-optical phase detector by amplitude modulation of the local field in a rydberg atom-based mixer,} {\protect\JournalTitle{Chinese Physics B}} \textbf{31}, 090703 (2022).

\bibitem{sedlacek2013PRL}
J.~Sedlacek, A.~Schwettmann, H.~K{\"u}bler, and J.~Shaffer, \enquote{Atom-based vector microwave electrometry using rubidium rydberg atoms in a vapor cell,} {\protect\JournalTitle{Physical review letters}} \textbf{111}, 063001 (2013).

\bibitem{wang2023OE}
Y.~Wang, F.~Jia, J.~Hao, \emph{et~al.}, \enquote{Precise measurement of microwave polarization using a rydberg atom-based mixer,} {\protect\JournalTitle{Optics Express}} \textbf{31}, 10449--10457 (2023).

\bibitem{Liu2023AtomVectorMicrowaveElectricField}
X.~Liu, F.~Jia, F.~Zhou, \emph{et~al.}, \enquote{Atomic vector microwave electric field meter based on electromagnetically induced transparency and autler-townes splitting of cold rydberg atoms,} {\protect\JournalTitle{Journal of Astronautic Metrology and Measurement}} \textbf{43}, 5--10 (2023).

\bibitem{li2023OE38165--38178}
X.~Li, Y.~Cui, J.~Hao, \emph{et~al.}, \enquote{Magnetic-field-induced splitting of rydberg electromagnetically induced transparency and autler-townes spectra in 87 rb vapor cell,} {\protect\JournalTitle{Optics Express}} \textbf{31}, 38165--38178 (2023).

\bibitem{shi2023OE36255--36262}
Y.~Shi, C.~Li, K.~Ouyang, \emph{et~al.}, \enquote{Tunable frequency of a microwave mixed receiver based on rydberg atoms under the zeeman effect,} {\protect\JournalTitle{Optics Express}} \textbf{31}, 36255--36262 (2023).

\bibitem{schlossberger2024PRAA}
N.~Schlossberger, A.~P. Rotunno, A.~B. Artusio-Glimpse, \emph{et~al.}, \enquote{Zeeman-resolved autler-townes splitting in rydberg atoms with tunable resonances and a single transition dipole moment,} {\protect\JournalTitle{Physical Review A}} \textbf{109}, L021702 (2024).

\bibitem{schlossberger2024zeeman}
N.~Schlossberger, A.~P. Rotunno, A.~B. Artusio-Glimpse, \emph{et~al.}, \enquote{Zeeman-resolved autler-townes splitting in rydberg atoms with tunable resonances and a single transition dipole moment,} {\protect\JournalTitle{Physical Review A}} \textbf{109}, L021702 (2024).

\bibitem{jia2020AO}
F.~Jia, J.~Zhang, L.~Zhang, \emph{et~al.}, \enquote{Frequency stabilization method for transition to a rydberg state using zeeman modulation,} {\protect\JournalTitle{Applied Optics}} \textbf{59}, 2108--2113 (2020).

\bibitem{Elgee2024three-dimensional}
P.~K. Elgee, K.~C. Cox, J.~C. Hill, \emph{et~al.}, \enquote{Complete three-dimensional vector polarimetry with a rydberg atom rf electrometer,}  (2024).

\end{thebibliography}

\end{document}